\documentclass[twocolumn]{article}

\usepackage{amsmath, amsthm, amsfonts}
\usepackage{hyperref}
\usepackage{natbib}
\usepackage{graphicx}

\theoremstyle{definition}

\theoremstyle{remark}

\title{A Possible Large-scale Alignment of Galaxy Spin Directions -- Analysis of 10 Datasets from SDSS, Pan-STARRS, and HST}

\author{Lior Shamir \\
\small Kansas State University \\
\small Manhattan, KS 66506, USA
}

\date{}

\begin{document}
\maketitle

\begin{abstract}

Multiple observations made by several different telescopes have shown asymmetry between the number of spiral galaxies rotating in opposite directions in different parts of the sky. One of the immediate questions regarding the possible asymmetry of the spin directions is whether the distribution forms a cosmological-scale axis. This paper analyzes and compares 10 different datasets published in the past decade, collected by SDSS, Pan-STARRS, and Hubble Space Telescope. The datasets contain spiral galaxies separated by their spin direction, and the distribution can show dipole axes. The analysis shows that the directions of the most probable dipole axes are consistent in datasets that have similar average redshift, but different between datasets that have different average redshift. The analysis also shows that the location of the most probable axis correlates with the average redshift of the galaxies in the datasets. That is, the location of the most probable axis shifts when the redshift gets higher, and the correlation is statistically significant. This provides a certain indication of a drift in a possible axis formed by the distribution of galaxy spin directions, or a cosmological scale structure that peaks at a certain distance from Earth. 

\end{abstract}


\section{Introduction}
\label{introduction}

The availability of digital sky surveys enabled by robotic telescopes has provided new approaches to observe the Universe, and particularly the large scale structure. One of the consistent observations is the non-random distribution of the spin directions of spiral galaxies \citep{macgillivray1985anisotropy,longo2011detection,shamir2012handedness,shamir2013color,shamir2016asymmetry,shamir2019large,shamir2020patterns,lee2019mysterious,lee2019galaxy,shamir2020pasa}.  The profiles of asymmetry show certain evidence of possible existence of a cosmological-scale dipole axis \citep{shamir2020patterns,shamir2020pasa,shamir2021particles}. When the redshift range of the galaxies is similar across different datasets, different telescopes show similar positions of the most likely dipole axis \citep{shamir2020patterns,shamir2020pasa}. However, if a cosmological-scale structure centered around an axis exists, it is also possible that the peak of such cosmological-scale structure is not centered around Earth.

The contention that the Universe is oriented around a cosmological-scale axis has been proposed through observations of cosmic microwave background \citep{mariano2013cmb,land2005examination,ade2014planck,santos2015influence}, and coined ``Axis of Evil''. While the observations of the cosmological-scale axis are consistent across data from the Cosmic Background Explorer (COBE), Wilkinson Microwave Anisotropy Probe (WMAP) and Planck, the existence of such axis is still controversial, and it has also been proposed that it can be the result of statistical fluctuations \citep{rassat2014planck}, or a local feature related to the effect of the solar system \citep{hansen2012can}.

Spiral galaxies are unique astrophysical objects in the sense that their visual appearance depends on the perspective of the observer. Earlier attempts to study the distribution of the spin directions of spiral galaxies were based on small datasets \citep{macgillivray1985anisotropy}, manual annotation of galaxies made by undergraduate students \citep{longo2011detection}, or by volunteers accessing the galaxy images through an online service \citep{land2008galaxy,slosar2009galaxy}. Due to the large size of the datasets, human identification can be too slow to fully annotate the datasets collected by digital sky surveys. To handle a very high number of spiral galaxies, other attempts were based on automatic analysis of the galaxy images \citep{shamir2012handedness,shamir2013color,shamir2016asymmetry,shamir2019large,shamir2020patterns,shamir2020pasa}. The automatic analysis has the advantage of being able to process far larger datasets, while also using fully symmetric algorithms to ensure that no perceptual bias can affect the analysis \citep{shamir2021particles}. Results showed similar profiles of asymmetry in datasets acquired by Sloan Digital Sky Survey (SDSS), the Panoramic Survey Telescope and Rapid Response System (Pan-STARRS), and the Hubble Space Telescope \citep{shamir2020patterns,shamir2020pasa}, all showing similar profiles of asymmetry for when the redshift range of the galaxies are also similar.

Non-random distribution of galaxy spin directions have been observed in cosmic filaments, where galaxies that are part of the same filament are also aligned in their spin directions \citep{tempel2013evidence,tempel2013galaxy}, although that spin is different than the two-headed alignment, where the spins show no preference compared to the direction of the alignment. Alignment in spin directions has been associated with the large-scale structure, as shown by the distribution of quasars \citep{hutsemekers2014alignment} and spiral galaxies \citep{lee2019mysterious}. 

In addition to observational studies, simulations of dark matter have also shown correlation between spin and the large-scale structure \citep{zhang2009spin,libeskind2013velocity,libeskind2014universal}. The strength of the correlation has been associated to stellar mass and the color of the galaxies \citep{wang2018spin}. These links were associated with halo formation \citep{wang2017general}, leading to the contention that the spin in the halo progenitors is linked to the large-scale structure of the early universe \citep{wang2018build}. It should be mentioned that the spin direction of a galaxy might not be necessarily the same as the spin direction of the dark matter halo, as it has been proposed that a galaxy can also spin in a different direction than its host halo \citep{wang2018spin}.

The multiple studies with different probes, observations, and approaches that suggest a link between galaxy spin directions and the large-scale structure require the profiling and analysis of the relevant data. Acquired by powerful robotic telescopes with high throughput, data of very many galaxies can be used to study the distribution of galaxy spin directions in the local Universe. In this paper the relationship between the redshift range and the most probable axis is examined through several different datasets of spiral galaxies. The datasets were collected and analyzed in previous experiments, have different mean redshifts, and in some cases the analysis also shows certain differences in the location of the most probable axis. The mean redshift of these datasets ranges between $\sim$0.04 in dataset compiled using the Sloan Digital Sky Surveys (SDSS) to $\sim$0.6 in Hubble Space Telescope (HST). The broad redshift range can allow studying the existence of possible cosmological-scale alignments that their signal peaks within a certain distance from Earth.

\section{Data}
\label{data}

The data used in this study is based on several datasets of spiral galaxies sorted by their spin directions, and reported in the existing literature. The datasets were acquired using different telescopes, which are SDSS, HST, and the Panoramic Survey Telescope and Rapid Response System (Pan-STARRS). The analysis was done by fitting the asymmetry in the distribution of galaxy spin directions to cosine dependence, and identifying the direction of the most likely axis formed by the asymmetry. That was done by attempting to correlate the spin directions of the galaxies from all possible $(\alpha,\delta)$ combinations to identify the strength of possible cosine dependence, and the most probable axis from each dataset is reported in the corresponding relevant paper that describes the dataset and its analysis.

Cosine dependence between the spin direction of galaxies and their position in the sky can be conceptualized as a dipole axis reflected by a higher number of galaxies spinning clockwise in one side of the pole, and a higher number of galaxies spinning counterclockwise in the opposite end. Such axis can be identified computationally by assigning each galaxy with a value within the set $\{-1,1\}$, such that galaxies that spin clockwise are assigned with 1, and galaxies that spin counterclockwise are assigned with -1. Then, $\chi^2$ statistics is used to identify cosine dependence between the actual spin direction of the galaxy and its location in the sky. 

Such dipole axis is computed from each possible integer $(\alpha,\delta)$ combination. For each $(\alpha,\delta)$, the angular distance $\phi$ between the celestial coordinates of each galaxy in the dataset and $(\alpha,\delta)$ is computed. The  $\cos(\phi)$ of all galaxies in the dataset are then fitted into $d\cdot|\cos(\phi)|$, such that $d$ is a value within the set \{-1,1\} that corresponds to the actual spin direction of the galaxy. The $\chi^2$ is computed 1000 times such that in each run the galaxies are assigned with random spin directions, and the mean and standard deviation are computed for each integer $(\alpha,\delta)$ combination. The $\chi^2$ mean computed with the random spin directions is compared to the $\chi^2$ computed using the real spin directions. The $\sigma$ difference between the $\chi^2$ computed with the actual spin directions and the mean $\chi^2$ computed when using the random spin directions shows the likelihood of a dipole axis at $(\alpha,\delta)$. Once the likelihood of all $(\alpha,\delta)$ is computed, the most likely $(\alpha,\delta)$ can be identified. A detailed description of the determination of the cosine dependence can be found in \citep{shamir2012handedness,shamir2020patterns,shamir2021particles}.

The datasets and the papers that describe them are listed in Table~\ref{datasets}. The full description of the acquisition, preparation, and analysis of each dataset is specified in the corresponding relevant paper mentioned in the table. The table also shows the location of the most likely axis and the mean redshift of the objects in each dataset. For datasets 1, 4, 6, 7, 8, and 9 all objects had spectroscopic redshift, allowing a straightforward determination of the mean redshift in these datasets. In the case of Dataset 5 the mean redshift was determined using a subset of $\sim10^4$ objects in that dataset that had spectra. 

In the case of Dataset 3, the maximum redshift of the galaxies was 0.085, but the exact redshift distribution is not specified in the paper \citep{longo2011detection}. Therefore, the mean redshift was estimated by the mean redshift of the same redshift range $(z<0.085)$ in the SDSS Galaxy Zoo galaxies classified as unbiased ``superclean'' by their spin directions in \citep{land2008galaxy}, and to another dataset of SDSS galaxies with the same redshift range $(z<0.085)$ classified by their spin directions automatically \citep{shamir2020patterns}. In both cases the mean redshift of the galaxies was $\sim$0.05, and therefore 0.05 was estimated as the redshift range of the galaxies in that dataset. 

Dataset 2 is based on Pan-STARRS galaxies, which do not have spectra in the Pan-STARRS data release. Therefore, the mean redshift was determined by matching Pan-STARRS galaxies with galaxies that were also imaged by SDSS and had spectra. From the 38,998 Pan-STARRS galaxies in that dataset, 12,186 galaxies had spectra through SDSS \citep{shamir2020patterns}. The mean redshift of the galaxies in Dataset 2 was therefore determined based on the galaxies that had spectra in SDSS. Dataset 10 is the only dataset where photometric redshift was used. Although photometric redshift is inaccurate, due to the nature of the instrument it is clear that the HST galaxies are on average far more distant than the galaxies imaged by the Earth-based instruments.

\begin{table*}
\scriptsize
\begin{tabular}{|l|c|c|c|c|c|c|c|c|}
\hline
\# & Dataset       & Telescope	                     & \# galaxies      & z  &  RA             & Dec              & Likelihood \\ 
    &  reference   &                                         &                      &      & (degrees)  & (degrees)        &               \\ 
\hline
1 & \citep{shamir2020patterns}     & SDSS           & 38,998      &  0.04    &   229      &   -21       &   $<2\sigma$ \\ 
2 & \citep{shamir2020patterns}     & Pan-STARRS & 33,028     & 0.04     &   227      &      1        &  $<2\sigma$ \\ 
3 & \citep{longo2011detection}      & SDSS          & 15,158        &   0.05  &  217          & 32         &   5.15$\sigma$ \\ 
4 & \citep{shamir2016asymmetry}  & SDSS &  13,440              &   0.06   &    165   &   30            &   4.02$\sigma$ \\ 
5 & \citep{shamir2021particles}      & SDSS	&  77,840              &  0.07    &  165            &  40      &      2.56$\sigma$ \\ 
6 & \citep{land2008galaxy}            & SDSS &  10,720              &  0.07     &     160         &    15     &    $<2\sigma$  \\ 
7 & \citep{shamir2012handedness}  & SDSS & 126,501          & 0.08  & 132             &  32             &       4.27$\sigma$ \\ 
8 & \citep{shamir2020patterns}     & SDSS   & 63,693           &  0.12   &  69             &  56             &  4.63$\sigma$      \\ 
9 & \citep{shamir2020pasa}     & SDSS        & 15,863            & 0.25   &    71           &   61            &       7.38$\sigma$  \\ 
10 & \citep{shamir2020pasa}     & HST        & 8,690             &  0.58   &      78          &   47          &    2.8$\sigma$     \\ 
\hline
\end{tabular}
\caption{Datasets of spiral galaxies, the mean redshift, and the RA and Dec of the most probable axis formed by the asymmetry in the spin directions of the galaxies.}
\label{datasets}
\end{table*}

The datasets in Table~\ref{datasets} were sorted into galaxies with opposite spin directions using different methods. The galaxies in Dataset 6 were sorted by using a large number of over $10^5$ non-expert volunteers who classified the galaxies by their spin direction through a web-based interface. Since the work of such a large number of volunteers is difficult to control and manage, only galaxies in which 95\% of the annotations agreed were used. That provided a relatively clean dataset, although strong bias in the data was still observed \citep{land2008galaxy}.

Dataset 3 was annotated manually by five undergraduate students. Despite the much lower number of human annotators and far lower investment of human hours, the dataset produced more annotated galaxies than Dataset 6. That can be attributed to the better control of the annotation process, as the identity of the annotators was well-known, making it easier to control the quality of their work, compared to the anonymous volunteers who worked through a web-based interface \citep{land2008galaxy}.

Datasets 4 and 10 were annotated manually by the author, in a long labor-intensive process. To avoid human perception bias, half of the galaxies were mirrored randomly, ensuring that any human bias would impact galaxies with both spin directions equally. The process was also repeated after mirroring all galaxies to ensure that the annotation is consistent.

The manually annotated datasets 3, 4, and 6 are also the three smallest datasets in Table~\ref{datasets}. To classify the larger datasets, a computer algorithm was used. Because even a small bias in the algorithm can lead to statistically significant signal, the classification algorithm must be fully symmetric. While machine learning, and specifically deep learning, are commonly used for automatic astronomical image classification, these data-driven algorithms are based on complex non-intuitive rules, and work as a ``black box''. Because the rules used by machine learning systems are not clear, they can capture any piece of information that improves the classification, and can therefore lead to biases. Such biases can include the preparation of the training set with different types of galaxies. For instance, having galaxies of more common morphologies in the class of clockwise galaxies can lead to an excessive number of galaxies classified as clockwise. 

To avoid the possibility of systematic bias, the fully symmetric model-driven Ganalyzer algorithm was used \citep{shamir2011ganalyzer}. Ganalyzer works by first converting the galaxy image into its radial intensity plot. The radial intensity plot of the galaxy image is the transformation of the original image such that the value of the pixel $(x,y)$ in the radial intensity plot is the median value of the 5$\times$5 pixels around $(O_x+\sin(\theta) \cdot r,O_y-\cos(\theta)\cdot r)$ in the original galaxy image, where $\theta$ is the polar angle and {\it r} is the radial distance.

Then, a peak detection \citep{morhavc2000identification} is applied to each horizontal line in the radial intensity plot, and the peaks are identified. Because pixels on the arms of the galaxy are brighter than other pixels at the same distance from the center of the galaxy, the peaks in the radial intensity plot are expected to correspond to galaxy arms. If the arm is curved, the peaks are expected to exhibit a non-straight vertical line in the radial intensity plot. When applying a linear regression to the vertical line, the sign of the regression coefficient determines the direction of the curve of the arm, and consequently the direction of rotation of the arm.

It is clear that not all galaxies are spiral, and not all spiral galaxies provide sufficient details to identify their spin direction. To remove galaxies with no identifiable spin direction, only galaxies that have at least 30 identified peaks that feature a non-straight vertical line are considered as galaxies with identifiable spin direction, and all other galaxies are being rejected as unidentifiable. That provides a fully symmetric deterministic algorithm that avoids making forced classifications of galaxies that do not have an identifiable spin direction. More detailed information about Ganalyzer and experimental results can be found in \citep{shamir2011ganalyzer,shamir2012handedness,hoehn2014characteristics,dojcsak2014quantitative,shamir2021particles,shamir2020patterns,shamir2020pasa}.

\section{Analysis of the datasets}
\label{analysis}

\begin{figure}[h]
\centering
\includegraphics[scale=0.6]{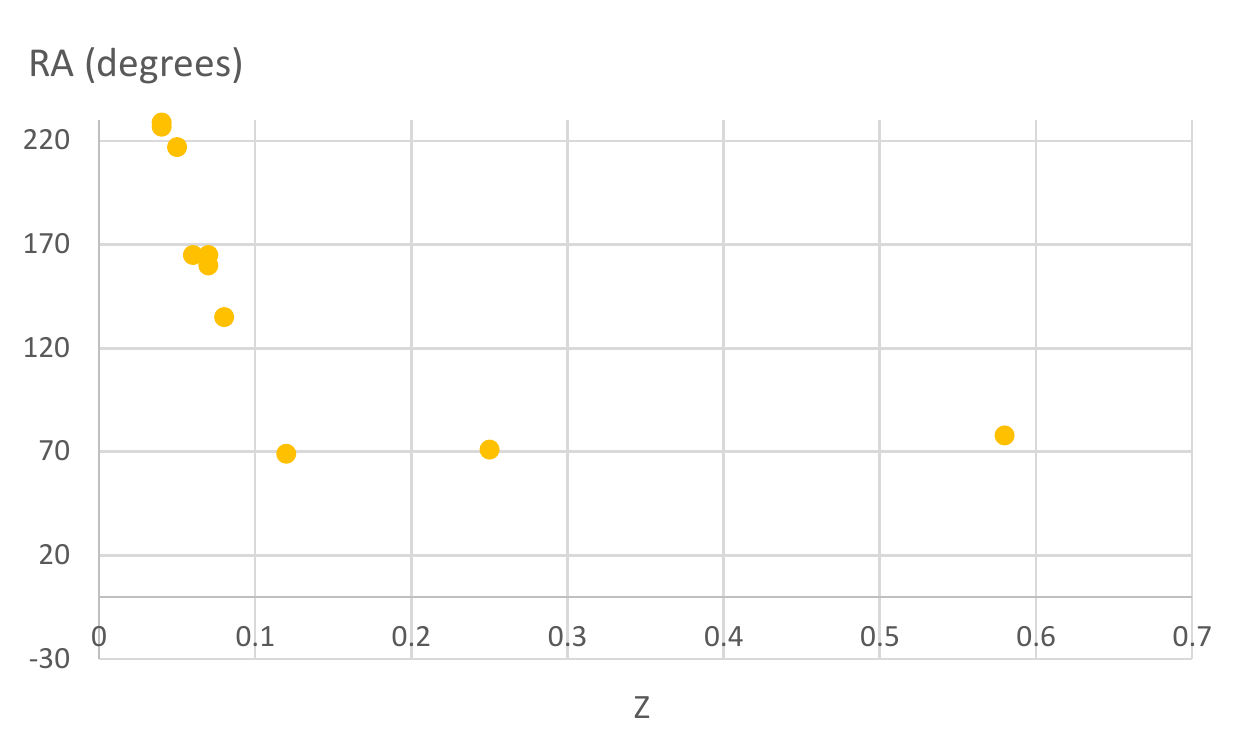}
\caption{The RA of the most probable axis identified in the different datasets. The x-axis shows the mean redshift of the galaxies in each dataset.}
\label{redshift_chart}
\end{figure}

The most likely dipole axes computed from the different datasets are shown in Table~\ref{datasets}. As the table shows, the right ascension of the most likely axis increases with the redshift. Figure~\ref{redshift_chart} displays the change in RA of the most probable axis with the redshift. The Pearson coefficient of the correlation between the RA and the redshift is -0.66. The one-tailed probability to have such correlation by chance is 0.0189, and the two-tailed probability is 0.038. That provides a certain indication of a possible correlation between the location of the most probable axis and the redshift of the galaxies in the dataset. The table also shows that when getting to redshift of 0.12 or higher, the RA of the most likely axis does not change significantly.

Assuming that the asymmetry between galaxies with opposite spin directions form an axis, the consistent change of the most likely axis with the redshift can indicate on a drift of the possible axis, if such axis indeed exists. 

\begin{figure}[h]
\centering
\includegraphics[scale=0.6]{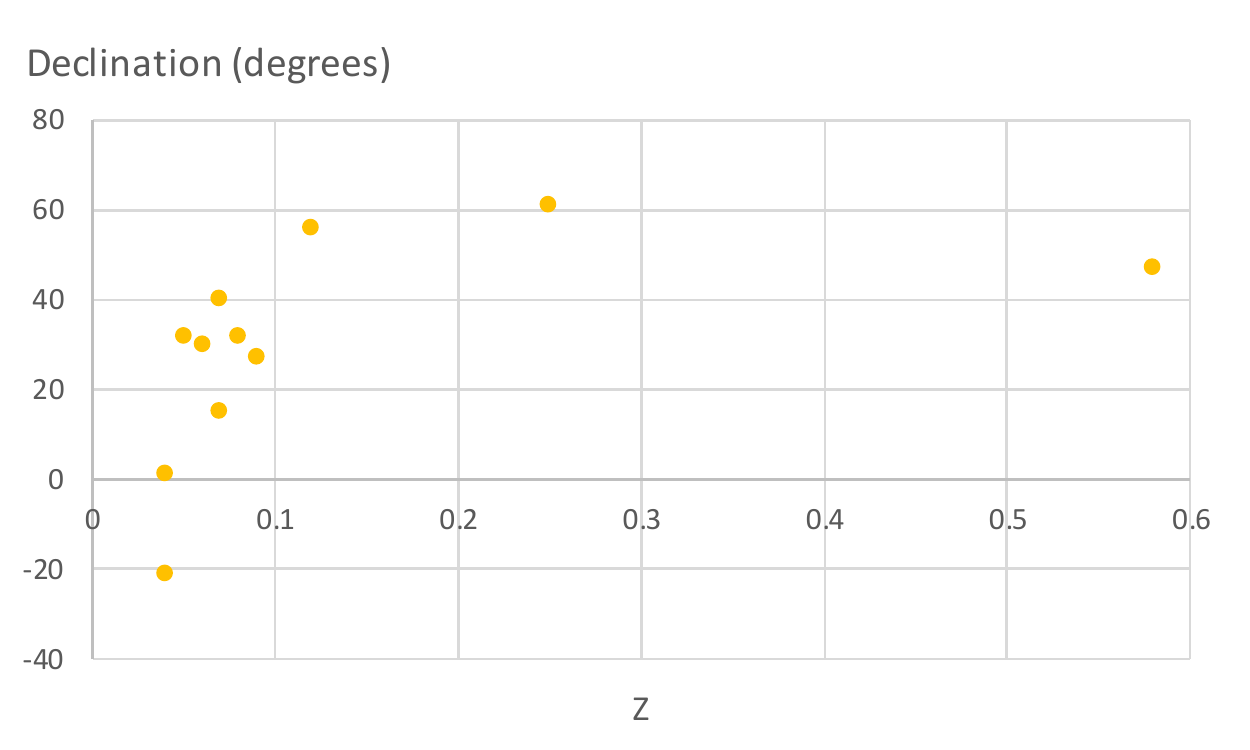}
\caption{The declination of the most probable dipole axis identified in the different datasets, and the mean redshift of each dataset.}
\label{declination_chart}
\end{figure}

Table~\ref{datasets} also shows certain evidence of a consistent change in the declination as the redshift gets higher. Figure~\ref{declination_chart} shows the declination of the most probable axis and the mean redshift of the galaxies in each dataset. The Pearson correlation between the z and the declination of the most probable axis is 0.48, with two-tailed P value of 0.16. While the correlation is not necessarily statistically significant, it shows a possible certain consistency between the redshift and the declination. It should be mentioned that the datasets are mostly taken from the Northern hemisphere, and the declination range is therefore relatively narrow, making it more difficult to determine the variations in the declination.

\begin{figure}[h]
\centering
\includegraphics[scale=0.22]{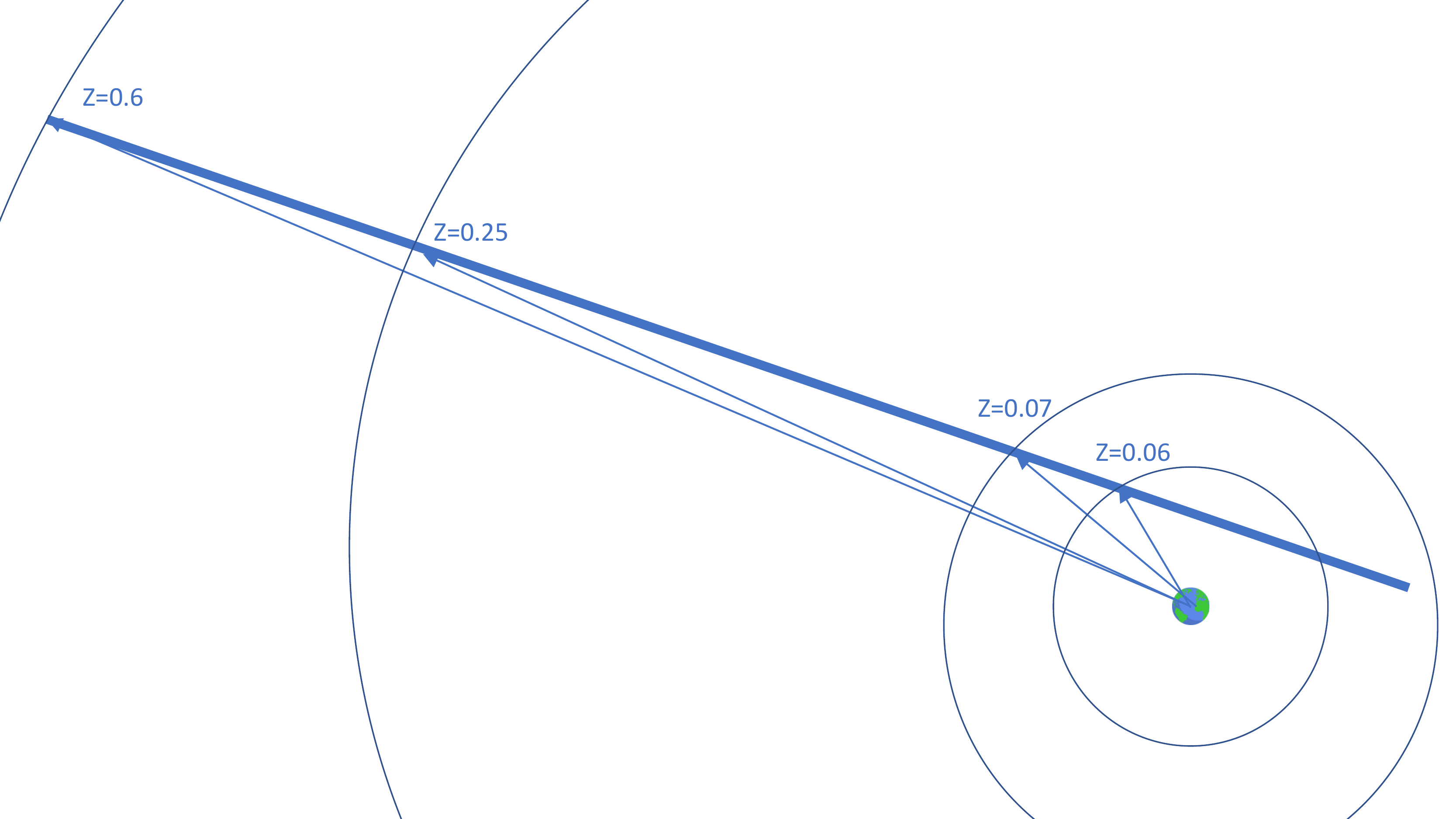}
\caption{Changes in the direction of a possible most likely axis as observed from Earth and changes with the redshift. The location of the peak of the axis does not change substantially for the higher redshifts, but the change is much larger when the redshift is lower.}
\label{restframe_axis3}
\end{figure}

The change in the right ascension of the location of the most probable axis as a function of the redshift can also indicate on an axis that peaks at a certain distance from Earth. That also agrees with the relatively small changes when the redshift is greater than 0.12. Figure~\ref{restframe_axis3} shows a simple illustration of how the right ascension changes when the redshift changes in an oversimplified universe of just two dimensions. The difference in the right ascension of the peak of the axis as observed from Earth changes with the redshift. It does not change substantially between the higher redshifts z=0.25 and z=0.6, while the difference in the RA is larger between z=0.06 and z=0.07.


Assuming the existence of a cosmological-scale axis, the similar locations of the axis at redshifts of 0.12, 00.25, and 0.58 as shown in Table~\ref{datasets} show the general direction of the possible axis as observed from Earth. The decreasing RA as the redshift gets higher can be used to estimate the distance between the peak of the most probable axis and Earth. Assuming a two-dimensional universe, and the direction of the most likely axis is 78$^o$, the distance between the peak of the most probable axis and Earth can be determined from each dataset by $D=sin(\phi-\Phi) \cdot d$, where $\phi$ is the most probable axis identified when the mean distance of the galaxies is $d$, and $\Phi$ is the general direction of the most probable axis, which can be estimated based on the dataset with the highest mean redshift, which is the HST dataset peaking at around $\alpha=78^o$. Since the transformation from redshift to distance is not linear, the distance $d$, and consequently $D$, is measured in Mpc.

The four datasets 4, 5, 6, and 7 in Table~\ref{datasets} provide a distance $D$ of 264, 306, 304, and 293 Mpc, respectively. These distances are relatively close to each other. Interestingly, they are also close in both direction and distance to Shapely \citep{bolejko2008great,bardelli1994study}, at around 262 Mpc from Earth \citep{colin2017high}. Datasets 1, 2, and 3 provide distances of 149, 151, and 139 Mpc. However, the mean distance of the galaxies in these datasets is also much lower than the closest distance measured using Datasets 4 through 7, and therefore it is possible that many of the galaxies in these datasets are too far from the peak of the assumed cosmological-scale structure to show a possible cosmological-scale axis. That also agrees with the much lower statistical signal observed with datasets 1 and 2, which is merely $\sim1.8\sigma$ \citep{shamir2020patterns}. Datasets 8 and 9 also provide a shorter derived distance of 96 Mpc and 146 Mpc, respectively. But due to the sharp angle, any small change in the most likely axis has a significant impact on the derived distance. For instance, a change of 5$^o$ in the RA of the most likely axis of Dataset 9 provides a derived distance of 237 Mpc, while in Dataset 5 a change of 5$^o$ leads to a small change of merely 2 Mpc in the derived distance.

\section{Conclusions}
\label{conclusion}

In the past decade, several different observations with datasets of spiral galaxies separated by their spin directions have shown asymmetry between the number of galaxies with opposite spin directions, and the asymmetry changes based on the direction of observation. Analysis of a possible dipole axis in the different datasets shows consistency between datasets with similar redshifts, and certain differences between datasets that are different in the redshift of their galaxies. Based on the datasets, it is also possible that the direction of the axis changes based on the redshift of the galaxies, as the RA of the most likely axis decreases consistently as the redshift gets higher. The change in the RA with the redshift could indicate on a certain drift of a possible cosmological-scale axis, but could also be the result of a cosmological-scale structure reflected by the direction towards which galaxies spin. A three dimensional analysis shows a peak that passes through the approximate direction and distance of Shapely.

The existence of a cosmological-scale structure reflected by galaxy spin directions also agrees with smaller scale experiments \citep{lee2019mysterious}. While the size of the dataset is too small to identify a cosmological-scale structure, the data shows correlation between the spin directions of spiral galaxies even when the galaxies are too far to have gravitational interactions, and that correlation is defined as  ``mysterious'' \citep{lee2019mysterious}. These observations also conflict with the standard cosmology, unless the assumption of Newtonian gravity is expanded into modified Newtonian dynamics (MOND) models that can explain longer gravitational span \citep{amendola2020measuring}, while also explaining other anomalies in the context of the standard model such as the Keenan–Barger–Cowie (KBC) void \citep{haslbauer2020kbc}.

Alignment has been found in position angle of radio galaxies, suggesting large-scale consistency in their angular momentum \citep{taylor2016alignments}, and these observations have been consistent also in other datasets such as the Faint Images of the Radio Sky at Twenty-centimetres (FIRST) and TIFR GMRT Sky Survey (TGSS), showing large-scale alignment \citep{contigiani2017radio,panwar2020alignment}.

The asymmetry between galaxies with opposite spin directions forming a dipole axis is naturally difficult to explain by the standard cosmological theories. The large-scale asymmetry provides indication of cosmological-scale anisotropy, which has been reported also by using other probes
such as the cosmic microwave background \citep{eriksen2004asymmetries}, Ia supernova \citep{javanmardi2015probing,lin2016significance}, frequency of galaxy morphology types \citep{javanmardi2017anisotropy}, short gamma ray bursts \citep{meszaros2019oppositeness}, LX-T scaling \citep{migkas2020probing}, and quasars \citep{quasars}. Another notable observation that violates the assumption of cosmological isotropy is the statistically significant CMB cold spot \citep{cruz655non,mackenzie2017evidence,farhang2021cmb}. These observations challenge the basic assumptions on which the standard models are based on.

Evidence of cosmological-scale anisotropy and cosmological-scale dipole axes observed in the distribution of the cosmic microwave background \citep{cline2003does,gordon2004low,zhe2015quadrupole,luongo2021larger} has led to alternative cosmological theories. These include double inflation \citep{feng2003double}, contraction prior to inflation \citep{piao2004suppressing}, primordial anisotropic vacuum pressure \citep{rodrigues2008anisotropic}, multiple vacua \citep{piao2005possible}, moving dark energy \citep{jimenez2007cosmology}, and spinor-driven inflation \citep{bohmer2008cmb}. Understanding the nature and patterns of a possible cosmological-scale anisotropy can provide further information that can lead to more alternative theories or expansion of existing thoeries.

Early analyses of the dipole anisotropy of the cosmic microwave background showed that the dipole peaks at ($\alpha=195^o,\delta=32^o$) \citep{conklin1969velocity}, ($\alpha=171^o,\delta=3^o$) \citep{fabbri1980measurement}, or at ($\alpha=174^o,\delta=-12^o$) \citep{boughn1981dipole}, close to the datasets shown in Table~\ref{datasets}. Alignment between the CMB dipole anisotropy and the dipole exhibited by galaxy spin directions can indicate that the asymmetry in galaxy spin directions is primordial.


The alignment in spin directions between galaxies too far from each other to have a gravitational link can also be considered an indication of the external field effect arising from large-scale structure in the context of Milgromian dynamics \citep{Milgrom1983modification,famaey2012modified,banik2021from}.
Such gravitational dynamics can also be aligned with the Hubble tension by means of our location within a particularly wide and deep supervoid that is not expected in $\Lambda$CDM. While not expected by the standard model, such a void could arise in a Milgromian cosmology. That is, the expansion rate history, cosmic microwave background anisotropies, and primordial light element abundances would remain the same as in $\Lambda$CDM, but structure formation would be enhanced due to the modified gravity law. Enhanced structure formation compared to $\Lambda$CDM is also suggested by the El Gordo (ACT-CL J0102-4915) massive galaxy cluster, which has observed properties that conflict with $\Lambda$CDM at probability of 6.16 $\sigma$ \citep{asencio2020massive}. Other $\Lambda$CDM tensions focus on smaller scale observations, normally at scales smaller than 1 Mpc \citep{del2017small,bullock2017small}. For instance, the satellite plane and the number of dwarf satellite galaxies in the Local Group shifts from what $\Lambda$CDM predicts, and galaxies dominated by dark matter are not as dense as can be deduced from $\Lambda$CDM. Dwarf galaxies can also have a much flatter core than predicted by $\Lambda$CDM \citep{pozo2020detection}.

A possible cosmological-scale axis can also be related to theories such as rotating universe \citep{godel1949example,ozsvath1962finite,ozsvath2001approaches,sivaram2012primordial,chechin2016rotation,camp2021}. It can also agree with theories of ellipsoidal universe \citep{campanelli2006ellipsoidal,campanelli2007cosmic,gruppuso2007complete,campanelli2011cosmic,cea2014ellipsoidal}, where a cosmological scale symmetry axis is assumed, and has been associated with the axis formed by CMB anisotropy \citep{campanelli2007cosmic}.

More recent theories such as holographic big bang \citep{pourhasan2014out,altamirano2017cosmological} can also be related to a dipole axis in the spin direction of galaxies. An axis of cosmological scale is aligned with the theory of black hole cosmology \citep{pathria1972universe,easson2001universe,chakrabarty2020toy}, and provides an alternative explanation to cosmic inflation. Since stars normally spin, black holes also spin \citep{mcclintock2006spin}. Due to the spin of the host black hole, a cosmological-scale axis is expected in the universe it contains. If the Universe was born from a black hole, it should have a preferred direction inherited from the direction of the spin of its host black hole \citep{poplawski2010cosmology,seshavatharam2010physics,seshavatharam2020light}, an axis \citep{seshavatharam2020integrated}, and the observed black hole universe might not be aligned with the cosmological principle \citep{stuckey1994observable}.




The possibility of non-random distribution of the spin directions of spiral galaxies is a domain of research that became possible in the information era in astronomy research, driven by robotic telescopes that enable comprehensive digital sky surveys. The large collections of extra galactic objects provide new ways of observing the universe that were not possible in the pre-information era. It is clear that more research will be needed to profile the large-scale distribution of the spin directions of spiral galaxies, and fully understand the cosmological aspects of possible structures made by the distribution of spiral galaxies.

\section*{Acknowledgments}

This study was supported in part by NSF grants AST-1903823 and IIS-1546079. I would like to thank the two knowledgeable anonymous reviewers for the insightful comments, and specifically for identifying patterns that I did not notice initially in the data.


\end{document}